\documentclass[aps,pra,twocolumn,groupedaddress]{revtex4-1}
\usepackage{graphicx}

\begin{document}

\title{Low-temperature phase transitions in some quaternary solid solutions\\
of IV-VI semiconductors}

\author{A. I. Lebedev}
\email[]{swan@scon155.phys.msu.ru}
\author{I. A. Sluchinskaya}
\affiliation{Physics Department, Moscow State University, 119991 Moscow, Russia}


\begin{abstract}
Samples of PbS$_x$Se$_y$Te$_{1-x-y}$, Pb$_{1-x}$Sn$_x$Te$_{1-y}$Se$_y$, and
Pb$_{1-x}$Sn$_x$Te$_{1-y}$S$_y$ quaternary solid solutions were investigated
in the 4--200~K temperature range using electrical and X-ray methods. The
regions where low-temperature phase transitions occur were established.
It was shown that phase transitions in these solid solutions are associated
with off-center S and Sn ions. The dependence of the phase transition
temperature on the composition of solid solutions can be qualitatively
described taking into account the influence of substitutional disorder on
the ordering and tunneling of off-center ions.

\texttt{DOI: 10.1016/0925-8388(94)90713-7} [Journal of Alloys and Compounds {\bf 203},
51--54 (1994)]

\end{abstract}


\maketitle

\section{Introduction}

IV-VI family narrow-gap semiconductors are widely used for optoelectronic and
thermoelectric device fabrication. Although ternary solid solutions are used to
tune the spectral characteristics of infrared (IR) lasers and photodetectors,
higher performance of these devices can be achieved on heterostructures, which
use both ternary and quaternary solid solutions.

The low-temperature ferroelectric phase transitions are interesting features of
some IV-VI semiconductors. Such phase transitions were observed in binary GeTe
and SnTe compounds, in Pb$_{1-x}$Sn$_x$Te, Sn$_{1-x}$Ge$_x$Te, Pb$_{1-x}$Ge$_x$Te,
and PbTe$_{1-x}$S$_x$ ternary and some quaternary solid solutions~\cite{Jantsch99,
JPhysSocJap.49A.725,JETPLett.40.998,IzvAkadNaukFiz.51.1683}. The changes in the
physical properties of these semiconductors induced by phase transition may be
interesting for some applications. In particular, the weak temperature dependence
of the energy gap near the phase transition temperature $T_c$ may be used for
the fabrication of continuously tunable lasers~\cite{ApplPhysLett.21.505}, and
the strong reduction of tunneling currents in $p$--$n$ junctions near $T_c$ may be
used to improve the characteristics of photovoltaic detectors~\cite{SovPhysSemicond.12.671}.
As the temperature of the phase transitions (usually 20--200~K) lies within the
working temperatures of IR devices, and there is not much information on
low-temperature phase transitions in quaternary solid solutions, it was important
to study the low-temperature phase diagrams of quaternary solid solutions in order
to find out the regions where the phase transitions can affect the physical
properties of solid solutions.

\section{Experimental details}

Studies of three quaternary solid solutions (Pb$_{1-x}$Sn$_x$Te$_{1-y}$Se$_y$,
Pb$_{1-x}$Sn$_x$Te$_{1-y}$S$_y$, and PbS$_x$Se$_y$Te$_{1-x-y}$ were performed on
the samples within the whole region of the stability of the room-temperature
cubic phase. Single crystals as well as polycrystalline samples of $n$- and
$p$-type were studied. Single crystals were grown by the sublimation method.
Polycrystalline samples were prepared by alloying the appropriate amounts of
binary compounds or ternary solid solutions in evacuated silica ampoules with
subsequent annealing at 600--710$^\circ$C for 25--220~h. After annealing, the
samples were cooled in air or quenched in cold water to prevent the decomposition
of the solid solution. In order to obtain samples of $p$- or $n$-type conductivity,
a small amount of excess Pb or chalcogen was added to the alloys. The homogeneity
of the samples was checked by the X-ray method. Most of the investigation was
performed on polycrystals in order to know the exact sample composition.

\begin{figure*}
\includegraphics{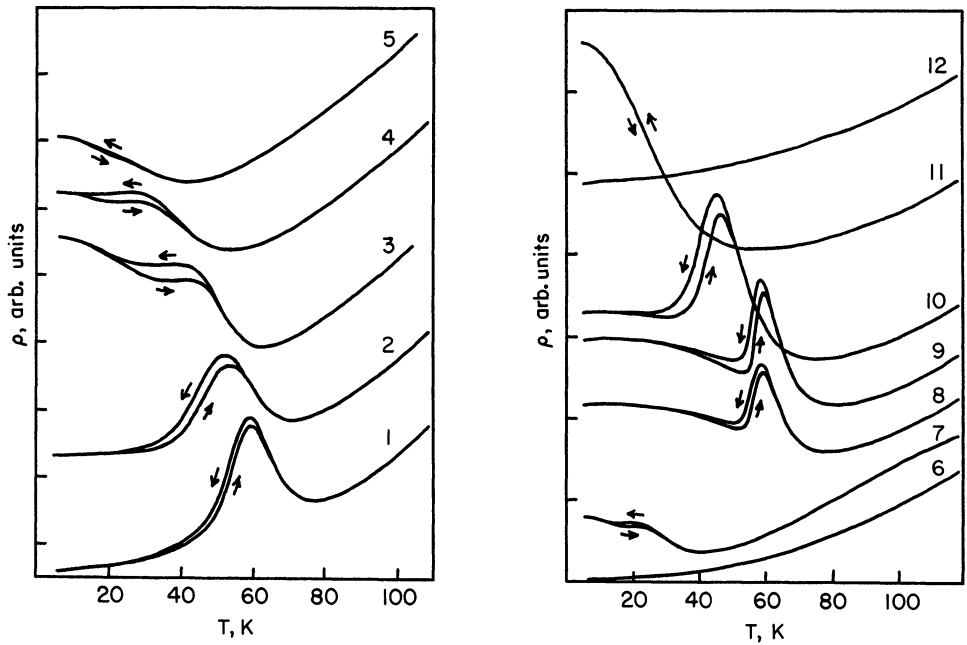}
\caption{\label{fig1}Typical $\rho(T)$ plots for PbS$_x$Se$_y$Te$_{1-x-y}$ and
Pb$_{1-x}$Sn$_x$Te$_{1-y}$Se$_y$ solid solutions. Curves 1--5 correspond to the
samples of PbS$_x$Se$_y$Te$_{1-x-y}$ with constant $x = 0.08$ and $y = 0.21$,
0.27, 0.34, 0.40, and 0.46, respectively. Curves 6--12 correspond to the samples
of Pb$_{1-x}$Sn$_x$Te$_{1-y}$Se$_y$ with constant $x = 0.2$ and $y = 0$, 0.05,
0.25, 0.33, 0.5, 0.67, and 0.83, respectively. The curves are shifted arbitrarily
along the vertical axis. The arrows show the direction of the temperature change
during recording the curves.}
\end{figure*}

The temperature dependence of resistivity $\rho(T)$ for these crystals was studied in
the range 4.2--200 K~\cite{SovPhysSolidState.25.2055}. Fig.~\ref{fig1} shows typical
$\rho(T)$ curves for PbS$_x$Se$_y$Te$_{1-x-y}$ and Pb$_{1-x}$Sn$_x$Te$_{1-y}$Se$_y$
samples. The anomalous resistivity peaks associated with the scattering of free
carriers by the ferroelectric fluctuations near the phase
transition~\cite{SolidStateCommun.17.875} are seen on most of the plots. The $\rho(T)$
curves obtained during heating and cooling (the directions of temperature change are
shown by arrows in Fig.~\ref{fig1}) differ in a wide temperature range around an
anomalous resistivity peak. The difference between the peak positions on heating
and cooling was usually 1--3~K, so the phase transition temperature $T_c$ was taken
as the mean value of the anomalous resistivity peak temperatures. At low temperatures
some $\rho(T)$ curves exhibited the enhancement of $\rho$ with decreasing temperature,
so-called additional low-temperature scattering. The causes of the appearance of
hysteresis and additional low-temperature scattering accompanying the phase transition
in crystals were discussed in detail in our previous
papers~\cite{IzvAkadNaukFiz.51.1683,JETPLett.46.536,SovPhysSolidState.32.1036,SovPhysSolidState.34.793}.

The phase transitions in Pb$_{1-x}$Sn$_x$Te$_y$Se$_{1-y}$ and Pb$_{1-x}$Sn$_x$Te$_{1-y}$S$_y$
crystals were studied by a low-temperature X-ray technique, which revealed the rhombohedral
distortion of the structure below $T_c$.

\begin{figure}
\includegraphics{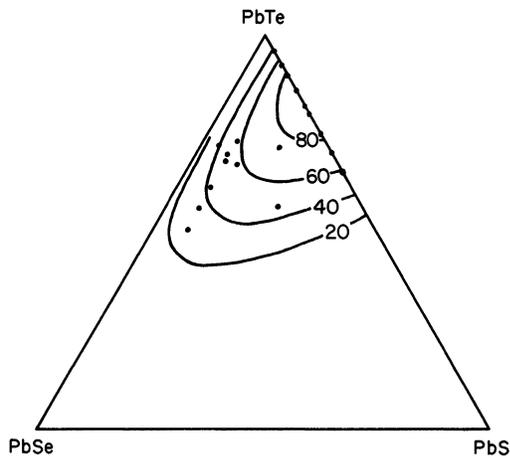}
\caption{\label{fig2}The dependence of $T_c$ (in K) upon composition of the
PbS$_x$Se$_y$Te$_{1-x-y}$ solid solution.}
\end{figure}

Figures \ref{fig2}--\ref{fig4} show the dependence of $T_c$ on the composition of
three quaternary systems (Pb--S--Se--Te, Pb--Sn--Te--Se and Pb--Sn--Te--S). The
points in these figures correspond to compositions of samples. The solid curves
are isotherms drawn after computer approximation of the experimental data.

\section{Discussion}

Our results show that in all three systems there are regions where the low-temperature
phase transitions occur. The manifestations of the phase transition in the electrical
properties of these solid solutions were similar to that in ternary Pb$_{1-x}$Ge$_x$Te
and PbTe$_{1-x}$S$_x$ solid solutions, where the phase transitions are unambiguously
associated with the presence of off-center ions~\cite{PhysRevLett.59.2701,Ferroelectrics.120.23}.
This enabled us to suppose that phase transitions in investigated quaternary solid
solutions have the same origin and are induced by off-center ions. Following the
results of Ref.~\onlinecite{JETPLett.40.998}, one could suppose that the S atom is
off-center in PbS$_x$Se$_y$Te$_{1-x-y}$, while according to
Refs.~\onlinecite{SovPhysSolidState.32.1036,SovPhysSolidState.34.793}, Sn is
off-center in Pb$_{1-x}$Sn$_x$Te$_{1-y}$Se$_y$, and Sn and S are both off-center
in Pb$_{1-x}$Sn$_x$Te$_{1-y}$S$_y$.

\begin{figure}
\includegraphics{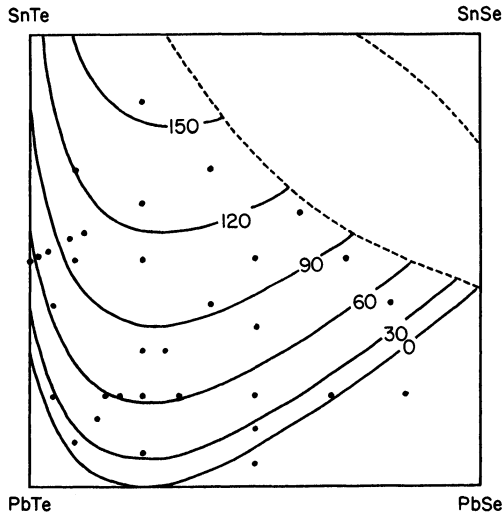}
\caption{\label{fig3}The dependence of $T_c$ (in K) upon composition of the
Pb$_{1-x}$Sn$_x$Te$_{1-y}$Se$_y$ solid solution. The dotted lines show the stability
limits for the solid solution.}
\end{figure}

\begin{figure}
\includegraphics{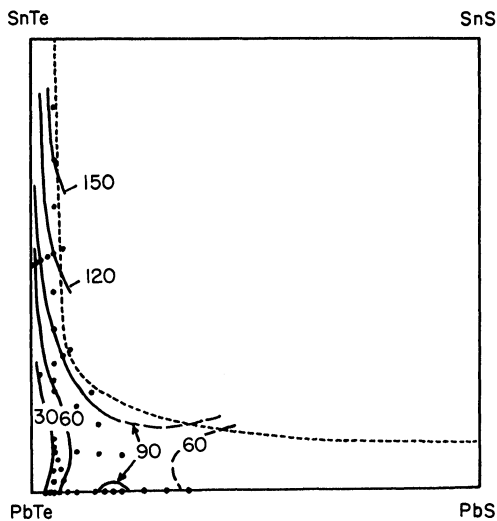}
\caption{\label{fig4}The dependence of $T_c$ (in K) upon composition of the
Pb$_{1-x}$Sn$_x$Te$_{1-y}$S$_y$ solid solution. The dotted line shows the stability limits
for the solid solution according our data.}
\end{figure}

The individual features of off-center ions and host crystals, into which off-center ions
are incorporated, result in a different dependency of $T_c$ on composition. Fig.~\ref{fig2}
shows that
maximum values of $T_c$ on the PbS$_x$Se$_y$Te$_{1-x-y}$ phase diagram are achieved in
ternary PbTe$_{1-x}$S$_x$ solid solutions; in other ternary solid solutions (PbS$_{1-x}$Se$_x$
and PbTe$_{1-x}$Se$_x$) there are no phase transitions. This emphasizes the key role of
substitution of Te by S in the appearance of phase transitions in PbS$_x$Se$_y$Te$_{1-x-y}$.
The region where phase transitions take place is located near PbTe whose lattice is known
to be softest among the different lead chalcogenides. Adding Se to PbTe$_{1-x}$S$_x$ lowers
$T_c$; one of the causes of this lowering may be the influence of random fields (resulting
from the substitutional disorder in the quaternary solid solution) on the ordering of
off-center ions~\cite{IzvAkadNaukFiz.51.1683}.

A qualitatively different behavior was observed in Pb$_{1-x}$Sn$_x$Te$_{1-y}$Se$_y$
solid solution. As follows from the phase diagram shown in Fig.~\ref{fig3}, the
substitution of Te by Se in Pb$_{1-x}$Sn$_x$Te crystals with low tin concentration
($x < 0.35$) results in the onset of the phase transition. When increasing Se
concentration in crystals with constant $x$, the phase transition temperature first
increased quickly, then reached a maximum (at about $y = 0.25$), and decreased again
down to zero when approaching Pb$_{1-x}$Sn$_x$Se. At constant $y = 0.25$, the phase
transition could be observed starting from Sn concentrations as low as $x = 0.08$.
According to our computer approximation of the experimental data, the maximum value
of $T_c$ should be achieved near $x = 1$ and $y = 0.25$, but another phase transition
took place in SnTe$_{0.75}$Se$_{0.25}$ at a higher temperature~\cite{SovPhysSolidState.32.1036}.
The maximum value of $T_c$ in this quaternary solid solution was 1.5~times higher
than in the best single crystals of SnTe~\cite{PhysRevLett.37.772}. The unusual rise
of $T_c$ with increasing the substitutional disorder in Pb$_{1-x}$Sn$_x$Te$_{1-y}$Se$_y$
was explained in Ref.~\onlinecite{SovPhysSolidState.32.1036}: the substitutonal
disorder does not only produce random fields, but it can depress strongly the tunneling
motion of off-center ions and so may result in an increase of $T_c$. We suppose that
the latter effect is predominant in Pb$_{1-x}$Sn$_x$Te$_{1-y}$Se$_y$.

Pb$_{1-x}$Sn$_x$Te$_{1-y}$S$_y$ is the most interesting system studied in this work.
There are two different off-center ions in this system, S and Sn. As follows from
Fig.~\ref{fig4}, their simultaneous effect on $T_c$ is more complicated than in
Pb$_{1-x}$Sn$_x$Te$_{1-y}$Se$_y$. The substitution of small amounts of Pb by Sn in
crystals with fixed S concentration first strongly decreased $T_c$ so that the phase
transition disappeared at about $x \approx 0.1$. With further increasing Sn concentration,
the phase transition appeared again. Thus, the low-temperature phase diagram consists
of two regions; in one of these (at low Sn concentration) the ferroelectric phase is
induced by S off-center ions, and in the other it is associated with Sn off-center
ions. The phase diagram is so complicated (it even has a saddle point) because the
substitutional disorder depresses the ordering of S off-center ions (by random fields)
and at the same time enhances the cooperative motion of Sn off-center ions due to
decreasing their tunneling.

The comparison of phase diagrams in Figs.~\ref{fig3} and \ref{fig4} shows that they
are similar in the region of high Sn concentration. The substitution of Te by S
results in a stronger increase of $T_c$ than in the case of substituting Te by Se;
we suppose that it is due to a stronger perturbation produced in the former case.
The similarity between these
phase diagrams confirms that the origin of the rise of $T_c$ in both cases is the same---the
depression of the tunneling motion of Sn off-center ions by the substitutional disorder~\cite{SovPhysSolidState.34.793}.
Unfortunately, the composition stability limits for Pb$_{1-x}$Sn$_x$Te$_{1-y}$S$_y$ are not
very large; the boundaries of the single phase region (according to our data) are indicated
by the dotted line in Fig.~\ref{fig4}.

\section{Conclusions}

1. The samples of PbS$_x$Se$_y$Te$_{1-x-y}$, Pb$_{1-x}$Sn$_x$Te$_{1-y}$Se$_y$, and
Pb$_{1-x}$Sn$_x$Te$_{1-y}$S$_y$ quaternary solid solutions have been studied at
low temperatures using electrical and X-ray methods. The regions where the
low-temperature phase transitions take place were established.

2. It was shown that the low-temperature phase transitions in all three systems are
associated with off-center S and Sn ions.

3. The dependence of the phase transition temperature on the composition of solid solutions
may be qualitatively described by taking into account the influence of substitutional disorder
on the ordering and tunneling of off-center ions.

\providecommand{\BIBYu}{Yu}

\end{document}